\documentclass[journal]{IEEEtran}
\ifCLASSINFOpdf
  \usepackage[pdftex]{graphicx}
  \usepackage{multirow}
\else
\fi
\usepackage{url}
\usepackage{xcolor}

\begin{document}
%
% paper title
% Titles are generally capitalized except for words such as a, an, and, as,
% at, but, by, for, in, nor, of, on, or, the, to and up, which are usually
% not capitalized unless they are the first or last word of the title.
% Linebreaks \\ can be used within to get better formatting as desired.
% Do not put math or special symbols in the title.
\title{Predicting popularity of online videos\\using Support Vector Regression}
%
%
% author names and IEEE memberships
% note positions of commas and nonbreaking spaces ( ~ ) LaTeX will not break
% a structure at a ~ so this keeps an author's name from being broken across
% two lines.
% use \thanks{} to gain access to the first footnote area
% a separate \thanks must be used for each paragraph as LaTeX2e's \thanks
% was not built to handle multiple paragraphs
%

\author{Tomasz~Trzci\'{n}ski and Przemys\l{}aw Rokita% <-this % stops a space
\thanks{T. Trzci\'{n}ski and P. Rokita are with the Institute of Computer Science, Warsaw University of Technology, ul. Nowowiejska 15/19, 00-665 Warsaw, Poland e-mail: t.trzcinski@ii.pw.edu.pl.}% <-this % stops a space
\thanks{Manuscript accepted.}}

% note the % following the last \IEEEmembership and also \thanks - 
% these prevent an unwanted space from occurring between the last author name
% and the end of the author line. i.e., if you had this:
% 
% \author{....lastname \thanks{...} \thanks{...} }
%                     ^------------^------------^----Do not want these spaces!
%
% a space would be appended to the last name and could cause every name on that
% line to be shifted left slightly. This is one of those "LaTeX things". For
% instance, "\textbf{A} \textbf{B}" will typeset as "A B" not "AB". To get
% "AB" then you have to do: "\textbf{A}\textbf{B}"
% \thanks is no different in this regard, so shield the last } of each \thanks
% that ends a line with a % and do not let a space in before the next \thanks.
% Spaces after \IEEEmembership other than the last one are OK (and needed) as
% you are supposed to have spaces between the names. For what it is worth,
% this is a minor point as most people would not even notice if the said evil
% space somehow managed to creep in.

% The paper headers
\markboth{Accepted to IEEE Transactions on 
Multimedia}%
{T. Trzci\'{n}ski, P. Rokita: Predicting popularity of online videos using Support Vector Regression}
% The only time the second header will appear is for the odd numbered pages
% after the title page when using the twoside option.
% 
% *** Note that you probably will NOT want to include the author's ***
% *** name in the headers of peer review papers.                   ***
% You can use \ifCLASSOPTIONpeerreview for conditional compilation here if
% you desire.

% If you want to put a publisher's ID mark on the page you can do it like
% this:
%\IEEEpubid{0000--0000/00\$00.00~\copyright~2015 IEEE}
% Remember, if you use this you must call \IEEEpubidadjcol in the second
% column for its text to clear the IEEEpubid mark.

% use for special paper notices
%\IEEEspecialpapernotice{(Invited Paper)}

\renewcommand{\topfraction}{1}
\renewcommand{\dbltopfraction}{1}
\renewcommand{\bottomfraction}{1}
\renewcommand{\textfraction}{.0}
\renewcommand{\floatpagefraction}{1}
\renewcommand{\dblfloatpagefraction}{1}

% make the title area
\maketitle

% As a general rule, do not put math, special symbols or citations
% in the abstract or keywords.
\begin{abstract}

In this work, we propose a regression method to predict the popularity of an online video measured by its number of views. Our method uses Support Vector Regression with Gaussian Radial Basis Functions. We show that predicting popularity patterns with this approach provides more precise and more stable prediction results, mainly thanks to the non-linear character of the proposed method as well as its robustness. We prove the superiority of our method against the state of the art using datasets containing almost 24,000 videos from YouTube and Facebook. We also show that using visual features, such as the outputs of deep neural networks or scene dynamics' metrics, can be useful for popularity prediction before content publication. Furthermore, we show that popularity prediction accuracy can be improved by combining early distribution patterns with social and visual features { and that social features represent a much stronger signal in terms of video popularity prediction than the visual ones}.
\end{abstract}

% Note that keywords are not normally used for peerreview papers.

% For peer review papers, you can put extra information on the cover
% page as needed:
% \ifCLASSOPTIONpeerreview
% \begin{center} \bfseries EDICS Category: 3-BBND \end{center}
% \fi
%
% For peerreview papers, this IEEEtran command inserts a page break and
% creates the second title. It will be ignored for other modes.
\IEEEpeerreviewmaketitle

\section{Introduction}
\label{sec:introduction}
\IEEEPARstart{R}{ecent} years have brought an enormous increase in the popularity of online platforms, such as YouTube, Facebook, Twitter or Instagram, where users can easily share various content with other people. YouTube is the biggest video sharing website with over 1 billion users that watch hundreds of millions of hours and generate billions of views~\cite{Youtube15}. The most popular social network with almost 1.5 billion registered users is Facebook~\cite{Facebook15}, followed by Instagram with over 400 million users~\cite{Instagram15} and Twitter with over 300 million active users sending 500 million tweets (short messages) per day~\cite{Twitter15}. Although not every social network user is equally active in creating and publishing content, it is estimated that 85\% of Facebook users actually do engage in the content creation process~\cite{Adage15}. Among different types of content generated by the users, photos and videos become more and more popular, mainly thanks to the proliferation of mobile devices with embedded high-quality cameras, but also as a result of studies indicating that visual content leads to higher user engagement~\cite{Twitter15a}. Since the amount of visual content accessible online is so high, one should expect that only a small portion of this data gains significant popularity, while the rest remains seen only by a small audience~\cite{Cha07}. This phenomenon has led to the inception of the term {\it viral video} which describes a movie uploaded online that is gaining audience in an exponential manner, often reaching millions of views within a few days of publishing.

In this context, the ability to predict the number of views of a given video can serve multiple causes, from load balancing the throughput of the data centers and servers to adjusting marketing efforts of the media houses that publish advertisements online. The latter application becomes increasingly significant, as marketing agencies spend 13\% more money on digital marketing each year, with an estimated \$52.8 billion spent in 2015~\cite{TechCrunch15}. { A typical approach to optimize those spendings is to use A/B testing of the content and adjust the content served to the consumers accordingly. Netflix reported that using A/B testing of the thumbnail images of the videos can lead up to a 30\% increase in video view counts~\cite{Netflix16}. Similar increase was also reported for A/B testing of opening video scenes on Facebook~\cite{Intelligence16}. This method, however, requires proper space sampling and can easily be biased if the selected group of testers is not large enough.} Moreover, social networks such as Facebook, allow the marketing agencies to promote their content by increasing the reach of their videos. In this context, estimating the future popularity of a video can improve the allocation of the promotional funds. For instance, if a video of a given publisher is expected to reach 1 million organic views and its predicted view count exceeds this number, the promotional funds can be spent on other less popular videos instead.

Predicting the popularity of videos published online is a challenging problem. First of all, the external context of the content plays an important role in the distribution patterns of the video, {\it i.e.} if the subject of a video is trending in other media (television, radio, newspapers), its popularity online is also expected to be high. Secondly, the structure of the network built around the publisher such as the number of its friends and followers, and their respective friends and followers, has a substantial impact on the distribution of the content and therefore its future popularity. Last but not least, factors such as the relevance of the video to the final viewer and the relationship between real world events and the content are complex and difficult to capture, increasing the difficulty of popularity prediction.

Nevertheless, in the recent years several attempts have been made to address the problem of online content popularity prediction~\cite{Szabo10,Borghol11,Bandari12,Pinto13,Khosla14,Xu14}. Researchers analysed several types of online content, including news articles~\cite{Bandari12}, Twitter messages~\cite{Osborne11,Hong11}, images~\cite{Khosla14,Gelli15} and videos~\cite{Borghol11,Pinto13,Wu16b,Wu16}. Proposed prediction methods rely either on intrinsic features of the content, such as visual or textual cues~\cite{Bandari12,Khosla14,Gelli15}, or on social features describing the structure of the social network~\cite{Xu14} or on early distribution patterns~\cite{Szabo10,Pinto13}. To our knowledge, not too much attention was paid to the problem of combining different cues to predict the popularity of the online content in the context of videos.
 
In this work, we propose a regression method based on Support Vector Regression with Gaussian Radial Basis Functions to predict the popularity of online videos.  We use visual cues as video features that can be computed before the video is published as well as early popularity patterns of the video once it is released online, including view counts and social interactions' data. We evaluate our method on datasets containing almost 24,000 online videos uploaded to YouTube and Facebook. The contributions of this paper are the following:
\begin{itemize}
\item We introduce a new popularity prediction method, named Popularity-SVR, for online video content that relies on Support Vector Regression (SVR) with Gaussian Radial Basis Function (RBF) kernel and show that it outperforms the state of the art.
\item We show that results obtained relying only on the early distribution patterns as done in~\cite{Szabo10,Pinto13}, can be improved by adding visual and social features, such as number of faces shown throughout the video or the number of comments recorded for a video.
\item We collect and open to the public a new dataset of  over 1,800 online videos uploaded to the largest social network along with the corresponding temporal and visual features.
\end{itemize} 
%Moreover our results show that it is difficult to estimate popularity of the video before it is published, but the precision of the estimation grows substantially with a few hours after publication. Furthermore, we show that results obtained relying only on the early distribution patterns as done in~\cite{Szabo10,Pinto13}, can be improved by adding visual metadata, such as number of faces shown throughout the video or size of text displayed.

The remainder of this paper is organized in the following manner. In Section~\ref{sec:related_work} we give an overview of the state of the art. In Section~\ref{sec:features} we discuss the features used to predict the popularity of online videos using methods described in Section~\ref{sec:methods}. Section~\ref{sec:results} presents the results and we conclude this work in Section~\ref{sec:conclusions}.
 
\section{Related work}
\label{sec:related_work}

\begin{figure*}[t]
\centering
\includegraphics[scale=.38]{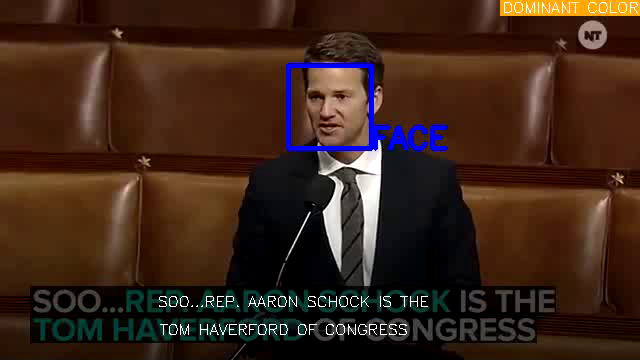}
\includegraphics[scale=.38]{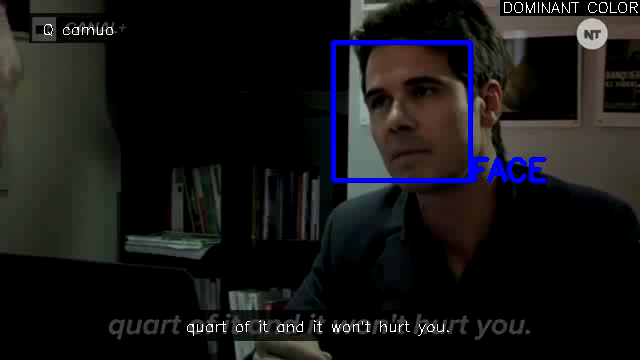}
\caption{Results of the visual content analysis of a sample video frame. The dominant color is displayed in the top right corner of the frame. The face is detected using cascade classifier. The text region is detected and faded to enable impainting OCR results. Best to be seen on a screen.}
\label{fig:visual_features}
\end{figure*}

Due to the enormous growth of the number of Internet users and online data available, popularity prediction of online content has received a lot of attention from the research community. Early works have focused on user web-access patterns~\cite{Almeida96} and more specifically on the distribution of the video content~\cite{Chesire01}, as it accounted for a significant portion of the Internet traffic and the findings could be used to determine the benefits of caching. Once the general access patterns were understood, the attention of the research community shifted to the actual popularity prediction of various content types.

Textual content, such as Twitter messages, Digg stories or online news, is typically distributed very fast and catches users' attention for a relatively short period of time~\cite{Castillo14}. Its popularity, measured in number of user actions such as comments, re-tweets or likes, is therefore highly skewed and can be modelled, e.g. with log-normal distribution~\cite{Tsagkias10}. Video content exhibits similar heavy-tailed distribution, while its popularity is typically measured by the number of views~\cite{Tatar14}.  The availability of the video content and related popularity data via the YouTube platform, where every minute over 100 hours of video is uploaded~\cite{Tatar14}, researchers were able to investigate other aspects related to the video content distribution. The most representative topics include prediction of the peak popularity time of the video~\cite{Jiang14} or identifying popularity evolution patterns~\cite{Crane08}. However, most if not all methods used to predict the popularity of a given video rely on its early evolution pattern~\cite{Szabo10,Borghol11,Pinto13} or its social context~\cite{Xu14}. Contrary to the method proposed in this paper, they do not exploit additional visual cues to improve their prediction accuracy.

In particular, Szabo and Huberman~\cite{Szabo10} observe a log-linear relationship between the views of the YouTube videos at early stages after the publication and later times. The reported Pearson correlation coefficient between the log-transformed number of views after seven and thirty days after publication exceeds 0.9, which suggests that the more popular submission is at the beginning, the more popular it will become later.

Building up on the log-linear model of~\cite{Szabo10}, \cite{Pinto13} proposed to extend their approach with Multivariate Linear (ML) model that uses multiple inputs from previous stages (values of views received by a video in the early times after publication) to predict the future popularity of the video. On top of the Ordinary Least Squares regressor, they also experimented with the Ridge regressor using Radial Basis Functions (RBF) which reduces the prediction error by 20\% on average with respect to the method of~\cite{Szabo10}. In this paper, we follow this lead and propose to use Gaussian RBF as a Support Vector Regression kernel~\cite{Drucker96}.

To improve the prediction accuracy, Xu et al.~\cite{Xu14} propose to add information about the structure of publisher's social network, including the proportion of the users who viewed and shared a video as well as the number of their followers. Their so-called Social-Forecast method aims to maximize the forecast reward defined as a trade-off between prediction accuracy and the timing of the prediction. Although the method shows improved accuracy in terms of forecast reward, it requires fairly detailed data concerning social network structure, which is not always available. For instance, Facebook, the social network with the highest number of registered users, does not allow to browse users' history of viewed videos and its followers' counts by public entities. Therefore, the Social-Forecast method, evaluated on the Chinese RenRen social network database where those metrics are publicly available, has to be adapted to other platforms if needed.

Although it is not the focus of this paper, a few approaches have been taken to predict the popularity of online content based on several information sources~\cite{Castillo14,Roy13}. For instance,~\cite{Roy13} use data from Twitter to detect YouTube videos that will receive a significant growth in popularity. The model is based on the extraction of popular and trending topics on Twitter and linking them to the corresponding YouTube videos. This results in 70\% higher accuracy of significant popularity growth prediction compared to the single-domain models that only use data from YouTube.

All the above mentioned works propose to predict future popularity of online content after the content is published. It is much more interesting, although more challenging as well, to attempt to predict the popularity of a given piece of content {\em before} it is published. Khosla et al.~\cite{Khosla14}  address this problem in the context of images. More precisely, the proposed method analyses visual and social features of the images published on Flickr to predict their relative popularity after the publication. Using a dataset of over 2 million images, the authors demonstrate that features such as image color or number of friends of the publisher play a significant role in determining the future popularity of a given photo. Moreover, using those cues, they are able to predict the normalized view count of images. {This work was later extended by Gelli et al.~\cite{Gelli15} to use visual sentiment and context features. 

Several recent works~\cite{Wu16,Wu16b} have also tackled the problem of image popularity in social media from a temporal perspective. Exploiting the popularity patterns and trends, Wu et al. proposed estimating popularity based on multi-scale analysis of the dependencies between user, time and item represented in Flickr pictures. 

We build on these works by proposing a popularity prediction method for social media videos}. We use computer vision algorithms to calculate visual features and verify if combining it with early evolution data can improve prediction accuracy for videos published online. {Although recent works have also addressed the problem of online video analysis~\cite{Zhang16} and popularity prediction~\cite{Chen16} from a multi-modal perspective, their focus is on micro-videos that last not more than a few seconds, while we consider longer videos}. To the best of our knowledge, this is one of the first attempts to use this kind of features in the context of online video popularity prediction.

% Our model incorporates the sequentiality of the information presented in the video, while~\citep{Chen16} analyses the frames separately. Finally, they rely on many other cues, such as textual and audio features, for the popularity prediction. In our work we use videos shared on Facebook that last much longer and, more importantly, we focus only on the visual cues that can be extracted from the videos. To our best knowledge, this is the first attempt to use only visual cues in the context of online video popularity prediction.  

\section{Features}
\label{sec:features}
In this section we discuss features of the videos used to predict their popularity. We start with the description of visual features that can be extracted before a video is published online. We then follow with an overview of temporal features recorded after the video was published. { In our terminology, the temporal features refer to the information that changes in a timely fashion, e.g. number of aggregated video views that increases with time or number of likes a given video receives that also changes in time}.

\subsection{Visual features}
Features presented here are computed using several computer vision algorithms applied on raw video data. The resulting features are then used to provide additional cues for the prediction methods.

{\bf Video characteristics:} We use simple video features describing video length, number of frames, video resolution and frame dimensions.

{\bf Color:} We first cluster the color space into 10 distinct classes depending on their coordinates in the Hue-Saturation-Value colorspace: {\it black, white, blue, cyan, green, yellow, orange, red, magenta} and {\it other}. Then, for each frame of a video, we assign a pixel to a single color and identify the dominant color of every frame. We aggregate the results of the color classification and represent color feature of a video as a histogram of dominant colors across the frames as well as dominant video color.

{\bf Face:} Using a face detector based on a cascade classifier~\cite{Viola01}, we detect the region of a frame with a face. We then count the number of detected faces per frame, number of frames with faces present and the size of the face regions with respect to the frame size. The results are averaged across all video frames and stored.

{\bf Text:} With a combination of edge detection and morphological filters, we identify the regions of the image with imprinted subtitles and apply Tesseract-OCR engine\footnote{\url{https://code.google.com/p/tesseract-ocr/}} to validate the detection. We then report the following textual characteristics of a video: a portion of the frames with imprinted text in the video and an average ratio of the text region size with respect to the frame size.

{\bf Scene dynamics:} To quantify scene dynamics of a video, we first employ Edge Change Ration algorithm~\cite{Jacobs04} and determine  shot boundaries. We then analyse the boundaries distribution and extract the number of shots and an average shot length in seconds. We also classify the shots as hard or soft cuts and save the corresponding histogram of shots.

{\bf Clutter:} We use a Canny edge detector~\cite{Canny86} to quantify the clutter present in the video. We report the ratio of the edge pixels detected and all pixels in a frame, averaged across all frames in a video.

{\bf Rigidity:} To evaluate the scene rigidity we estimate the homography between two consecutive frames using a combination of FAST feature point detector~\cite{Rosten10} and BRIEF descriptor~\cite{Calonder12}. We then save an average number of frames where a valid homography between current and previous frames can be found.

{\bf Thumbnail:} Building upon the work of~\cite{Khosla14}, we also compute a popularity score using Popularity API\footnote{\url{http://popularity.csail.mit.edu/}} of the video thumbnail and saved the result.

{{\bf Deep features}: To complement the set of visual features we use a recently proposed ResNet-152~\cite{He15} - a deep convolutional neural network with 152 layers which recently won the $1^{st}$ place in the ImageNet classification, detection and localization challenges. For each video, we first extract a set of thumbnails representing each scene. We propagate them through the ResNet-152 network and average the resulting 1000-dimensional probability output vector over all the thumbnails of a given video. Finally, we normalize the vector so that all its elements sum up to one.}

Fig.~\ref{fig:visual_features} shows a sample result of the computer vision analysis of two video frames.

\begin{figure}[t]
\centering
\includegraphics[width=0.49\textwidth]{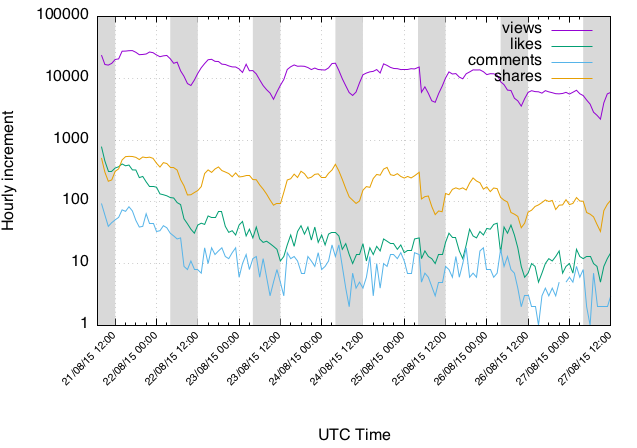}
\caption{Plot of hourly increments in number of views, likes, comments and shares for a sample Facebook video. The grey areas indicate night time according to the EDT Time Zone. The evolution patterns of those metrics are used in the paper to predict popularity of a given video.}
\label{fig:temporal_features}
\end{figure}

\subsection{Temporal features}
Once a video is made available online we are able to collect data related to its popularity that is the number of views as well as other social interactions aggregates. { We can therefore introduce the following features:
\begin{itemize}
\item {\bf Views}: an aggregated number of times a given video is watched that increases with time,
\item {\bf Social}: aggregated number of likes, shares and comments on a given video that also change in time.
\end{itemize}}
Figure~\ref{fig:temporal_features} shows a set of hourly increments in views, likes, comments and shares for a sample Facebook video. The evolution patterns of the video statistics provide an important cue for the popularity prediction methods, as~\cite{Szabo10} reported high correlation between log-transformed view counts early after the publication and later on. These results are also confirmed by the experiments presented in Section~\ref{sec:results}.

\section{Methods}
\label{sec:methods}
In this paper, following the works of~\cite{Szabo10,Pinto13} we cast the problem of popularity prediction as a regression task. More precisely, our goal is to predict the {\it number of views} of a video $v$ at time $t_t$, given features available from the first $t_r$ days after publication (where $t_r < t_t$). In this section, we discuss the regression methods used for the prediction in Section~\ref{sec:results}. We start by discussing the state-of-the-art methods in Section~\ref{subsec:state-of-the-art}. We then follow with the description of our proposed method called Popularity-SVR in Section~\ref{subsec:svr}.

\subsection{State-of-the-art methods}
\label{subsec:state-of-the-art}
{First, we discuss a set of state-of-the-art techniques, such as Univariate Linear (UL) Regression, Multivariate Linear (ML) Regression and Multivariate Radial Basis Function (MRBF) Regression.}

\subsection*{Univariate Linear (UL) Regression}
Based on the high correlation observed between log-transformed early and late popularity counts of online content,~\cite{Szabo10} proposed to use a simple regressor to predict the future popularity of a given video $v$. According to this model, the number of views of a video $v$ can be calculated at time $t_t$ as:
\begin{equation}
\hat{N}(v, t_r, t_t) = \exp \left( \alpha (t_r, t_t) \cdot \ln N(v, t_r) \right),
\end{equation}
where $\exp$ defines natural exponential function, $\hat{N}(v, t_r, t_t)$ defines predicted number of views for video $v$ at time $t_t$ when prediction is made at time $t_r$. $\alpha(t_r, t_t)$ is a weight learnt from training videos $v_t \in T$ and $N(v, t_r)$ is the number of views at time $t_r$. Weight $\alpha(t_r, t_r)$ can be computed using the ordinary least squares model.

\subsection*{Multivariate Linear (ML) Regression}
Pinto et al.~\cite{Pinto13} propose to extend the UL regression model by including also the views accumulated by the video before $t_r$. In other words, they increase the dimensionality of the input feature vector. Instead of using a single cumulated view count at time $t_r$, they sample the timeline between publication time $t_0$ and reference time $t_r$ and use the number of views received in those sampling intervals (views' {\it increments} or {\it deltas}) to form a feature vector. The proposed method called Multivariate Linear (ML) Regression predicts the popularity of the video $v$ at time $t_t$ as a linear combination of the feature values and can be expressed as:
\begin{equation}
\hat{N}(v, t_r, t_t) = \sum_{i=1}^{r} \alpha (t_i, t_t) \cdot \big( N(v, t_i) - N(v, t_{i-1}) \big),
\end{equation}
where $\{\alpha (t_i, t_t)\}_{i=1}^{r}$ are model parameters learned from training data $T$ and the term $\big( N(v, t_i) - N(v, t_{i-1}) \big)$ corresponds to the view deltas in the $i$-th sampling interval. 

\subsection*{MRBF Regression}
The ML Regression model is able to capture more information about the evolution pattern thanks to different weights assigned to time intervals. However, the weights learned from the training data cannot capture the intrinsic variations of the evolution patterns within the training dataset videos. Therefore,~\cite{Pinto13} propose to extend their ML model by introducing a similarity notion between the videos based on their evolution patterns. The so-called MRBF regression uses Radial Basis Functions (RBF) to calculate the distance between the videos and predicts the number of views based on the views increments as well as distances to a set of pre-selected training videos $v_c \in C$:
\begin{eqnarray}
\hat{N}(v, t_r, t_t) &=& \underbrace{\sum_{i=1}^{r} \alpha (t_i, t_t) \cdot \big( N(v, t_i) - N(v, t_{i-1}) \big)}_{\mbox{ML regression}} + \nonumber \\
 &+&  \underbrace{\sum_{v_c \in C} \omega_{v_c} \cdot \Phi(v, v_c)}_{\mbox{RBF features}},
 \label{eq:mrfb}
\end{eqnarray}
where $\Phi(x, y) = \exp \left(  - \frac{|| x - y ||^2}{2\sigma^2}\right)$ is a Gaussian RBF with $\sigma$ parameter and a set of videos $C$ to be selected during cross-validation.
The above problem can be solved with ordinary least squares, similarly to the previously discussed methods. However, the additional set of input features increases the risk of overfitting. Therefore,~\cite{Pinto13} propose to use Ridge regression~\cite{Hastie01} instead.

{It is worth mentioning that the MRBF regression uses Gaussian Radial Basis Function as a proxy for a similarity measure between the evolution patterns of a given video and a set of representative videos $v_c \in C$ from a training dataset. More precisely, the MRBF method postulates selecting a uniformly distributed random set of videos as representative samples. Then, the Gaussian RBF function is used to compute the distances between an input video and a set of samples. Finally, those distances are plugged into Eq.~\ref{eq:mrfb} and contribute to the RBF features' term of the prediction formula. This way the final popularity prediction of the MRBF method takes into account both the temporal popularity evolution of a given video (the ML regression term) and its similarity to previously observed popularity patterns within the training dataset (the RBF features' term).}

\subsection{Popularity-SVR}
\label{subsec:svr}
{MRBF Regression model encompasses linear and non-linear dependencies within the popularity evolution patterns using a combination of two methods: ML regression (linear) and RBF features (non-linear). This approach allows to compute the predicted value by combining the linear regression model based on the popularity evolution of a given video as well as its similarity to a set of representative videos from the dataset computed using a non-linear RBF kernel. We claim that it is not necessary to split the prediction into two distinct parts, which increases the complexity of the model and leads to additional computational costs. 

To this end, we propose a new method, dubbed Popularity-SVR, that predicts future popularity of a video using Support Vector Regression (SVR)~\cite{Drucker96}. { Inspired by the results obtained with the MRBF method, we propose to use Gaussian Radial Basis Functions as a kernel of our transformation. The selection of the right kernel can significantly influence the performance of the model, as it was shown in other domains, e.g. speaker identification~\cite{Boujelbene10} or handwriting recognition~\cite{Zhang07}. We therefore postulate using a RBF kernel as it allows us to map feature vectors into a non-linear space where the relations between popularity evolution patterns of the videos are easier to capture}. As a result, the non-linear character of the RBF kernel transformations allows for a more robust prediction based on the patterns identified by the algorithm within the training dataset, and not relying explicitly on the linear relation between early and later popularity of a given video. Therefore, Popularity-SVR simplifies the MRBF model by finding the relevant evolution patterns from within the training dataset and predicting the popularity based on the RBF-based similarity to those patterns. This approach is much different from the MRBF method, where the representative videos are selected as a uniform random sample of examples from the training dataset and the prediction is made based on the early evolution pattern and similarity to the random videos from the training dataset.} { Our approach also differs from similar works on modifying SVM kernel functions~\cite{Amari99,Boujelbene10,Zhang07}, since we consider using the RBF kernel as a method to generalize a more complex model proposed in~\cite{Pinto13}. Furthermore, to the best of our knowledge, our work is one of the first attempts to select an optimal SVM kernel in the context of online content popularity prediction.}

According to the proposed Popularity-SVR method the popularity of a video $v$ can be predicted as:
\begin{equation}
\hat{N}(v, t_r, t_t) = \sum_{k=1}^{K} \alpha_k \cdot \Phi \Big( X(v, t_r), X(k, t_r) \Big) + b,
\end{equation}
where $\Phi(x, y) = \exp \left(  - \frac{|| x - y ||^2}{2\sigma^2}\right)$ is a Gaussian RBF with $\sigma$ parameter, $X(v, t_r)$ is a feature vector for video $v$ available at time $t_r$ and $\{ X(k, t_r) \}_{k=1}^K$ is a set of support vectors returned by the SVR algorithm along with a set of coefficients $\{\alpha_k\}_{k=1}^K$ and intercept $b$. Unless stated otherwise, we use a vector of log-transformed view deltas as feature vectors, as proposed in~\cite{Pinto13}, that is $X(v, t_r) = \{ N(v,t_i) - N(v, t_i -1) \}_{i=1}^r$. We found optimal values for the hyperparameter $C$ of the Support Vector Machine optimization and $\sigma$ of the RBF kernel with a grid search in a preliminary set of experiments and in the remainder of this paper the following values are used: $C = 10, \sigma = 0.005$.

\setlength{\tabcolsep}{10pt}
\renewcommand{\arraystretch}{1.3}
\begin{table*}[t!]
\caption{\label{tab1}{YouTube video datasets. Results of the prediction for UL, ML, MRBF and Popularity-SVR methods reported as Spearman rank correlation $\pm$ 95\% confidence interval ($t_r = 6 \mbox{ days}, t_t = 30 \mbox{ days}$). Popularity-SVR outperforms the competitors while providing more stable prediction accuracy (smaller confidence interval).}}
\vspace{0.2cm}
\centering
\begin{tabular}{c|c|c|c||c}
\bf Dataset & UL & ML& MRBF& Popularity-SVR \\
\hline 
Random dataset &
0.8719 $\pm$ 0.0087 &
0.8844 $\pm$ 0.0087 & 
 0.8968 $\pm$ 0.0074 & 
\bf 0.9071 $\pm$ 0.0043 \\
\hline
 Top dataset &
 0.8797 $\pm$ 0.018 &
 0.8921 $\pm$ 0.017 &
 0.9046 $\pm$ 0.0152 &
 \bf 0.9353 $\pm$ 0.009
 \label{tab:flavio}
\end{tabular}

\end{table*}
\begin{figure*}[t!]
\centering
\includegraphics[width=0.49\textwidth]{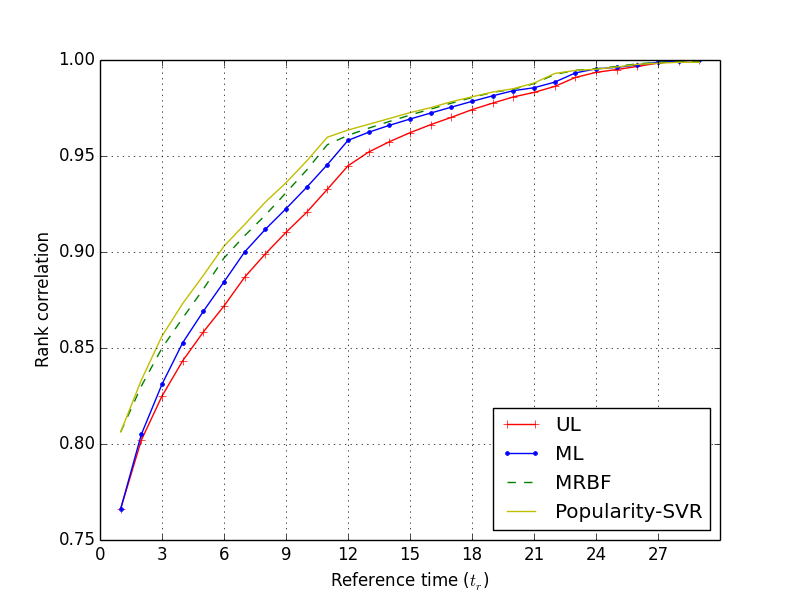}
\includegraphics[width=0.49\textwidth]{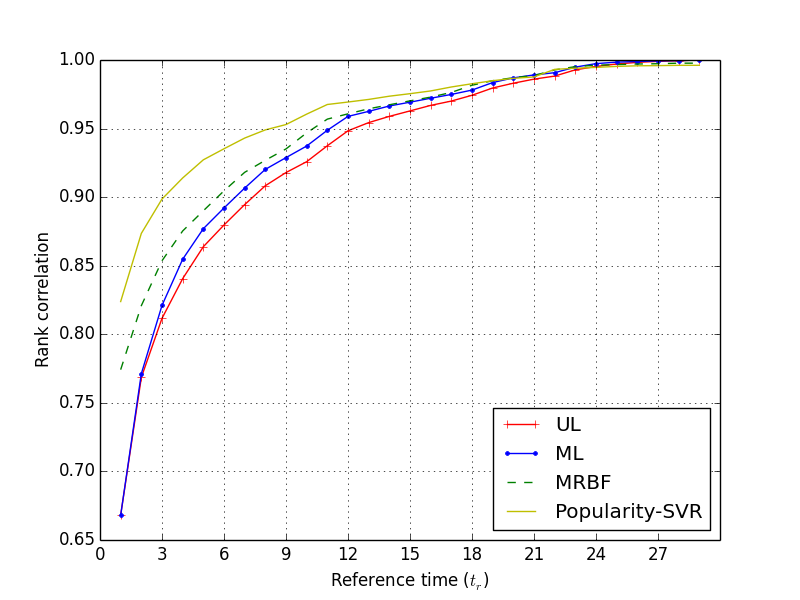}
\caption{Prediction results for the YouTube video datasets: Random (left) and Top (right). The reference time $t_r$ indicates number of days since publication and the target time is $t_t = 30$ days. The proposed Popularity-SVR method outperforms the state-of-the-art methods, among which the MRBF performs the best, for both datasets. The performance improvement is more significant for $t_r < 12$.}
\label{fig:flavio_results}
\end{figure*}

\section{Results}
\label{sec:results}
In this section we compare the state-of-the-art methods described in section~\ref{sec:methods}, namely the UL, ML and MRBF against the proposed Popularity-SVR method. To that end, we employ 3 datasets containing almost 24,000 videos. For ML and MRBF methods we use implementations obtained from their authors. For UL and Popularity-SVR we use our own Python implementation based on the Scikit-learn package\footnote{\url{http://scikit-learn.org/}}. {To find the optimal parameters of the MRBF and Popularity-SVR methods, we used Python scikit's {\tt sklearn.grid\_search.GridSearchCV }{ method, that finds the optimal parameters in terms of prediction accuracy using grid a search approach}, while the UL and ML implementations do not have any parameters to optimize}. We first evaluate all the methods using only the temporal evolution of the views (without visual or social features) on two publicly available datasets of YouTube videos: Top and Random~\cite{Figueiredo14}. We then show how we can improve the prediction precision with additional visual and social features obtained using a new dataset of Facebook videos.

\subsection{Datasets}
\label{subsec:datasets}
Top and Random datasets~\cite{Figueiredo14} contain data gathered for YouTube videos, such as time evolution of the number of views, comments, favorites and ratings. The Top dataset is a compilation of those results for a total of 27,212 videos taken from the top-100 most popular videos of each country in the world. The Random dataset contains the same type of data gathered for 24,484 unique randomly selected videos. {Similarly to~\cite{Pinto13}, we also preprocess both YouTube datasets and remove the videos with incomplete statistics and with less than 30 days of data. The final preprocessed datasets generated this way have 16,132 (Random) and 5,811 (Top) videos.}

To evaluate the prediction methods in the context of social media, we also collected data for 1,820 videos uploaded to Facebook between August $1^{{st}}$, 2015 until October $15^{th}$, 2015. The videos were uploaded by several Facebook publishers, including the AJ+\footnote{\url{www.facebook.com/ajplusenglish}} and BuzzFeedVideo\footnote{\url{www.facebook.com/BuzzFeedVideo}}. We implemented  a crawler that uses Facebook Graph API\footnote{\url{https://developers.facebook.com/docs/graph-api}} to browse Facebook publishers' pages and retrieve publicly available information regarding the number of interactions with a given video, that is the number of shares, likes and comments. Since the number of views of a video is not publicly available through the Graph API, we retrieve this data using simple URL scraper of a video page. { We release this dataset to the public to enable further research on the topic of popularity prediction of social media content\footnote{\url{http://ii.pw.edu.pl/~ttrzcins/facebook_dataset_2015.csv}}}.

\subsection{Evaluation protocol}
To evaluate the performance of prediction methods, we follow the approach of~\cite{Pinto13} and use 10-fold cross validation. For every dataset used, we randomly split all the samples into 10 equal-sized folds. We then use 9 folds for training and one for testing. We repeat the process 10 times, every time testing the methods on a distinct fold and training them with the remaining 9 folds. We report here the average results across all the 10 test sets along with the corresponding 95\% confidence interval. { As a metric to evaluate the prediction accuracy we use the Spearman rank correlation, as in~\cite{Khosla14}}.

\subsection{YouTube datasets}
\label{subsec:youtube_datasets}
We first evaluate the popularity prediction methods on two sets of YouTube videos: Random and Top datasets. Figure~\ref{fig:flavio_results} shows the results in terms of rank correlation for reference time $t_r \in (1,29)$ days and target time $t_t=30)$. The performance of our proposed Popularity-SVR method is higher than the competitors for both datasets and across the reference time values. The improvement over the state-of-the-art methods is more significant for the Top dataset and for $t_r < 12$, which indicates that our proposed method works especially well for the popular videos just after they are published. The performance of all methods converges as $t_r$ gets closer to the target time $t_t$. This is not a surprise, as the more time passes, the easier the prediction is. Out of the competitors, MRBF performs the best which confirms the results of~\cite{Pinto13}. For the quantitative analysis, we also show the average results along with the 95\% confidence interval for $t_r = 6$ in Table~\ref{tab:flavio}. {Not only does the Popularity-SVR method perform best, but its 95\% confidence interval is also up to 40\% smaller than the other methods, which means that Popularity-SVR provides a more stable prediction accuracy across different videos.}

\subsection{Facebook dataset}
{Secondly, we evaluate the performance using the Facebook dataset. In the first experiment we evaluated the Spearman rank correlation obtained when using various visual features proposed in Section~\ref{sec:features}. The results are shown in Table~\ref{tab:visual_features}. Our results show that the popularity of a video can be predicted with the highest accuracy using deep learning features, which confirms the observations made in~\cite{Khosla14} for images. Other important metrics that can be useful for the prediction of video popularity include clutter present in the video, scene dynamics and thumbnail popularity rank~\cite{Khosla14}. Interestingly, the negative correlation results obtained for text and rigidity features suggest that videos with too much text (e.g. subtitles) or those with too much rigidity are bound to be less popular, although the magnitude of the correlation is fairly small. Finally, combining all the visual features together provides the Spearman correlation result of over 0.23.}

\setlength{\tabcolsep}{3pt}
\renewcommand{\arraystretch}{1.4}
\begin{table}[t!]

\caption{\label{tab1}{Comparison of the video popularity prediction results using visual features. The results of the Popularity-SVR applied to groups of visual features proposed in Section~\ref{sec:features} show that deep features provide the {highest Spearman correlation value with video popularity. Overall correlation value using visual features reaches over 0.23 and is consistent with the results presented in~\cite{Khosla14} for images.}}}
\vspace{0.2cm}
\centering
\begin{tabular}{c||c}
{\bf Visual features} & {\bf Correlation}\\
\hline
Deep features & 0.1361 $\pm$ 0.0155 \\
Clutter & 0.1201 $\pm$ 0.0084 \\
Scene dynamics & 0.0822 $\pm$ 0.0091 \\
Thumbnail & 0.0682 $\pm$ 0.0094\\
Video characteristics & 0.0678 $\pm$ 0.0168\\ 
Face & 0.0588 $\pm$ 0.0188 \\
Color & 0.0385 $\pm$ 0.0103\\
Text & -0.0157 $\pm$ 0.0076\\
Rigidity & -0.0454 $\pm$ 0.0139 \\
\hline
Combined & {\bf 0.2344 $\pm$ 0.0166}
 \label{tab:visual_features}
\end{tabular}
\end{table}

\begin{figure}[t!]
\centering
\includegraphics[width=0.49\textwidth]{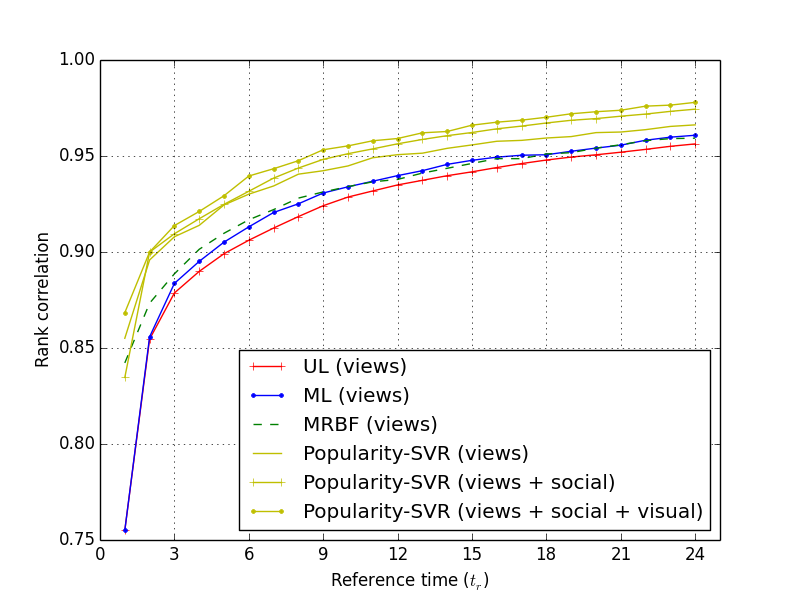}
\caption{Prediction results for Facebook dataset. The reference time $t_r$ indicates number of hours since publication and the target time is $t_t = 7$ days. Popularity-SVR provides better performance than other methods. When adding other types of data to the feature vector, the performance of Popularity-SVR is improved even more, reaching the peak with features based on the time evolution of the views, social and visual features.}
\label{fig:facebook_results}
\end{figure}

{We then compared the performance of the proposed Popularity-SVR method and state of the art using visual features, social features, such as the number of comments, likes and shares, and view counts as the inputs. Figure~\ref{fig:facebook_results} and Table~\ref{tab:facebook} show the obtained results. The results show that the highest prediction accuracy can be obtained using the combination of view counts, social features and visual features as inputs. When using those input sets separately, the best performance is observed for the view counts followed closely by social features. Although the correlation is not as high for the visual features, one must remember that visual features can be computed before the publication, while the others cannot be obtained until the video is published and it is too late to modify its contents. Therefore, we claim that the proposed visual features can be useful for the publishers to adjust the content and maximise its probability to become popular. 

{ Another conclusion we can draw from the results presented in Table~\ref{tab:facebook} is that social signals are much stronger in predicting popularity of online videos in social media than the visual signal. This confirms the findings of~\cite{Khosla14} for images shared online. One can consider the results of our experiment as an the empirical evidence that adding social features as an input of the online video popularity prediction methods leads to much higher improvement in terms of accuracy than adding visual features.}

The performance of all the methods can be improved by combining different feature subsets, although the improvement is modest as using only view counts provides a fairly high prediction accuracy. Moreover, the social features and view counts are highly correlated, as the more popular the content is, the more attention from other users it attracts. In fact, the Spearman correlation between the number of views and comments, shares and likes is equal to: 0.86, 0.88 and 0.93, respectively. This phenomenon, also known as multicollinearity, results in a relatively small information gain provided by those social features and explains the minor improvement over the views evolution data.}

\begin{figure}[t!]
\centering
\includegraphics[width=0.49\textwidth]{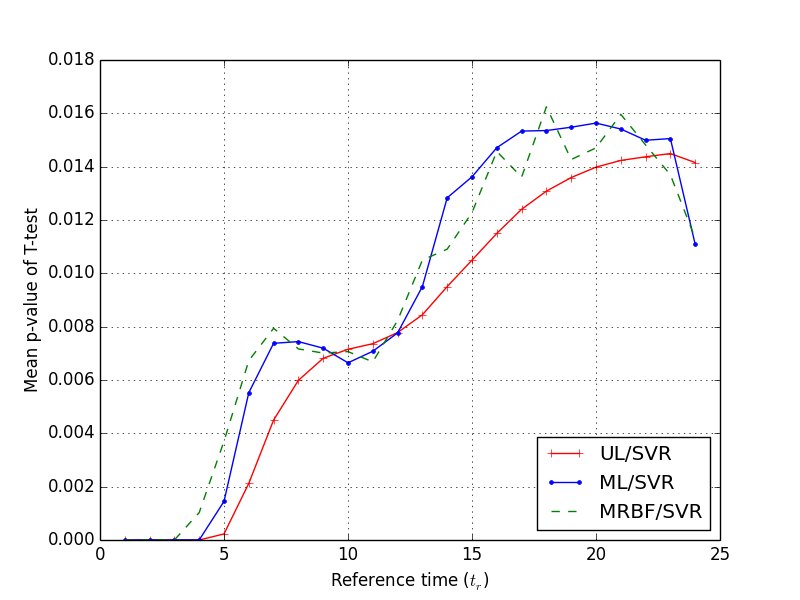}
\caption{ Results of Student T-tests in terms of p-values averaged over 10 test runs. The tests are run on the prediction results computed by the proposed Popularity-SVR method and the baseline methods on Facebook dataset. The statistical tests prove that the results are significantly different with mean p-value below 0.02 and, therefore, that the improvement of the proposed method over the state-of-the-art methods is statistically significant.}
\label{fig:t_test_results}
\end{figure}

Finally, the results confirm that our proposed Popularity-SVR method performs better than the competitors for all input configurations and across all $t_r$ values. {As the improvement of Popularity-SVR over MRBF for the input configuration with all features is approximately 1\%, we perform an additional set of experiments to verify the statistical significance of the results. To that end, we compute the Student-T test results for prediction outputs of { all baselines methods} and Popularity-SVR method using view counts. Figure~\ref{fig:t_test_results} shows the resulting p-values averaged across test folds. Although with the increasing time the results become more similar (with average p-value increasing), they remain statistically different with mean p-values below 0.02.}

%We also evaluated the performance of the Popularity-SVR when relying only on the visual features available before the publication of the video. As shown in Table~\ref{tab:facebook}, the average rank correlation in this case is equal to 0.197 which is much lower than the results obtained for the Popularity-SVR based on the temporal evolution of the views. Nevertheless, combining the data from both temporal (evolution of the number of views and social interactions) and visual features improves the overall prediction accuracy and provides the best performance among all evaluated methods, as can be seen in Figure~\ref{fig:facebook_results}.

\setlength{\tabcolsep}{3pt}
\renewcommand{\arraystretch}{1.4}
\begin{table*}[t!]

\caption{\label{tab1}{Facebook videos dataset. Results of the prediction for the UL, ML, MRBF and Popularity-SVR methods reported as Spearman rank correlation $\pm$ 95\% confidence interval ($t_r = 6 \mbox{ hours}, t_t = 7 \mbox{ days}$)}. Popularity-SVR method outperforms the state of the art methods across various input data configurations. Prediction accuracy of all the methods when using only visual features is lower than the accuracy of the methods relying on the temporal features. Nevertheless, combining visual and temporal features leads to higher accuracy. The best performance among all tested configurations is achieved by the Popularity-SVR method with time evolution of the number of videos, social interactions and visual features.}
\vspace{0.2cm}
\centering
\begin{tabular}{c||c|c|c||c}
& UL & ML & MRBF & Popularity-SVR\\
\hline
visual & 0.0957 $\pm$ 0.0184 & 0.1634 $\pm$ 0.0153 & 0.1496 $\pm$ 0.0133 & {\bf 0.2344 $\pm$ 0.0166} \\
social & 0.8658 $\pm$ 0.0386 & 0.8736 $\pm$ 0.0298 & 0.8728 $\pm$ 0.0305 & {\bf 0.8974 $\pm$ 0.0205} \\
views & 0.9061 $\pm$ 0.0366 & 0.9130 $\pm$ 0.0382 & 0.9173 $\pm$ 0.0379 & {\bf 0.9301 $\pm$ 0.0191} \\
views + visual & 0.9107 $\pm$ 0.0315 & 0.9152 $\pm$ 0.0301 & 0.9193 $\pm$ 0.0197 & {\bf 0.9311 $\pm$ 0.0125} \\
temporal (views + social) & 0.9126 $\pm$ 0.0295 & 0.9187 $\pm$ 0.0234  & 0.9197 $\pm$ 0.0237 &  {\bf 0.9356 $\pm$ 0.0160} \\
temporal (views + social) + visual & 0.9148 $\pm$ 0.032 & 0.925 $\pm$ 0.032 & 0.9203 $\pm$ 0.0366 & {\bf 0.9413 $\pm$ 0.0127}
 \label{tab:facebook}
\end{tabular}
\end{table*}

{
\subsection{Runtime evaluation}
For a novel prediction method to be used in practice, it needs to have low runtime (both in terms of training and prediction), as well as high scalability. To verify that our proposed approach fulfils this requirement, we measured execution times of all the methods for subsets of different sizes from the YouTube Random dataset presented in Section~\ref{subsec:datasets} and compared the results. The measurements were averaged over 10 runs and performed on a MacBook Pro with 2.5GHz Intel  Core i7 with 16GB RAM memory. Fig.~\ref{fig:runtime} shows the results of this comparison. Training time of our proposed SVR-Popularity method is lower than the other methods, except for the Univariate Linear (UL) Regression, while the prediction time is lower than the competing MRBF method. { We believe that it is the result of a simplified prediction model that uses only a set of support vectors along with the RBF kernel during prediction, while the MRBF method aditionally uses the ML regression term, as defined in Eq.~\ref{eq:mrfb}. Furthermore, our model can be trained faster than the competing MRBF approach, since it takes advantage of the so-called {\it kernel trick}~\cite{Hofmann09}. Employing the kernel trick allows us to avoid an explicit transformation of feature vectors into multi-dimensional RBF space and therefore reduces memory and computational costs}.  Moreover, increasing training size leads to increased training time for all the methods, while the prediction time remains fairly stable, proving the scalability of the evaluated methods.

\begin{figure*}[t!]
\centering
\includegraphics[width=0.49\textwidth]{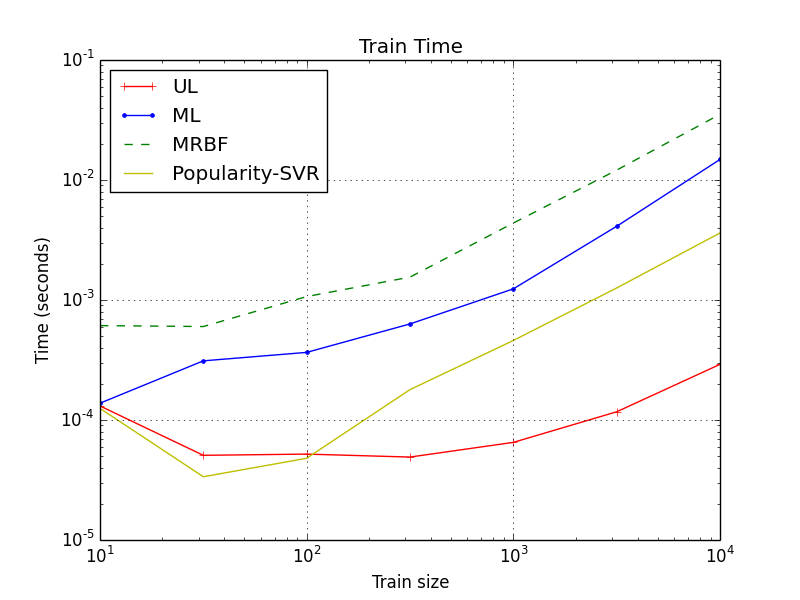}
\includegraphics[width=0.49\textwidth]{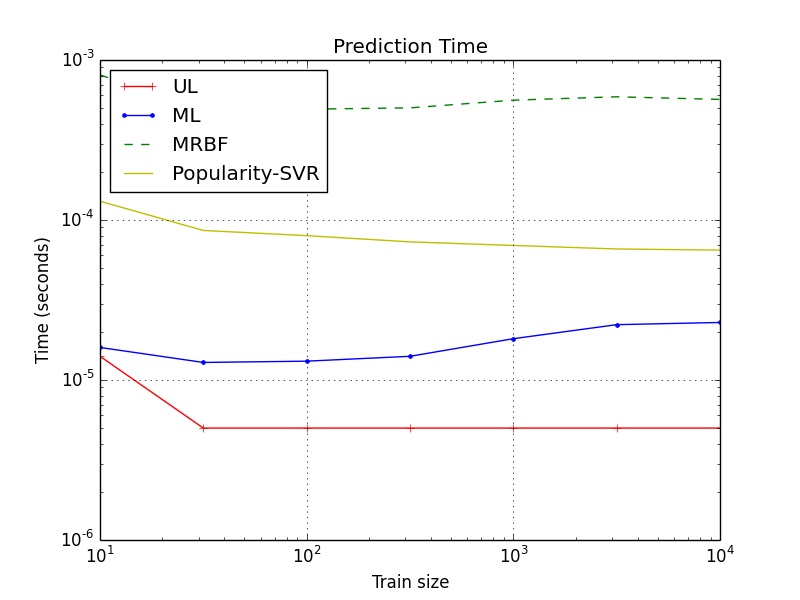}
\caption{Runtime evaluation comparison for various prediction methods on subsets of YouTube Random dataset of different sizes. Training time increases with the training set size, while prediction time remains stable across different sizes. The proposed Popularity-SVR method provides training times comparable to the state-of-the-art methods, while being faster at prediction than the competing MRBF approach.}
\label{fig:runtime}
\end{figure*}
}

\section{Conclusion}
\label{sec:conclusions}
In this paper, we propose to use Support Vector Regression with Gaussian Radial Basis Functions to predict the popularity of online video content measured as the number of views. 
%We showed that theing the dependencies between popularity evolution patterns with Gaussian RBF kernel leads to a significant performance improvement. 
{Our method was evaluated on three datasets containing a total of almost 24,000 videos and the results show its superiority with respect to the state of the art. Moreover, the results suggest that using only visual features computed before the publication of the video can be helpful to predict future video popularity. Nevertheless, if a higher prediction accuracy is required, temporal features, such as view counts or social features should be added. The best results obtained for the combination of visual features, social features and early view counts allow to predict the popularity of the { video published on Facebook} with a Spearman correlation rank of up to 0.94 only 6 hours after publication. In our future work we plan to extend the set of features used for prediction by adding more semantic cues, such as video topic or the sentiment of the social interactions, to better understand what impacts the popularity of the videos in social networks.}

\section*{Acknowledgments}
This work was partially funded by the grant of the Dean of the Faculty of Electronics and Information Technology at Warsaw University of Technology (project II/2015/GD/1).

%\section*{References}
\bibliographystyle{IEEEtran}

\begin{thebibliography}{10}
\providecommand{\url}[1]{#1}
\csname url@samestyle\endcsname
\providecommand{\newblock}{\relax}
\providecommand{\bibinfo}[2]{#2}
\providecommand{\BIBentrySTDinterwordspacing}{\spaceskip=0pt\relax}
\providecommand{\BIBentryALTinterwordstretchfactor}{4}
\providecommand{\BIBentryALTinterwordspacing}{\spaceskip=\fontdimen2\font plus
\BIBentryALTinterwordstretchfactor\fontdimen3\font minus
  \fontdimen4\font\relax}
\providecommand{\BIBforeignlanguage}[2]{{%
\expandafter\ifx\csname l@#1\endcsname\relax
\typeout{** WARNING: IEEEtran.bst: No hyphenation pattern has been}%
\typeout{** loaded for the language `#1'. Using the pattern for}%
\typeout{** the default language instead.}%
\else
\language=\csname l@#1\endcsname
\fi
#2}}
\providecommand{\BIBdecl}{\relax}
\BIBdecl

\bibitem{Youtube15}
YouTube, ``Press statistics,''
  \url{https://www.youtube.com/yt/press/statistics.html}, 2015, [Online;
  accessed 19-October-2015].

\bibitem{Facebook15}
Facebook, ``Company info,'' \url{http://newsroom.fb.com/company-info/}, 2015,
  [Online; accessed 06-October-2015].

\bibitem{Instagram15}
Instagram, ``Press,'' \url{https://instagram.com/press/}, 2015, [Online;
  accessed 06-October-2015].

\bibitem{Twitter15}
Twitter, ``Company info,'' \url{https://about.twitter.com/company}, 2015,
  [Online; accessed 06-October-2015].

\bibitem{Adage15}
Adage.com, ``Facebook 85 users creating content,''
  \url{http://adage.com/article/digital/facebook-85-users-creating-content/236358/},
  2015, [Online; accessed 06-October-2015].

\bibitem{Twitter15a}
Twitter, ``What fuels a tweet engagement,''
  \url{https://blog.twitter.com/2014/what-fuels-a-tweets-engagement/}, 2015,
  [Online; accessed 16-October-2015].

\bibitem{Cha07}
M.~Cha, H.~Kwak, P.~Rodriguez, Y.~Ahn, and S.~Moon, ``{I tube, you tube,
  everybody tubes: analyzing the world's largest user generated content video
  system},'' in \emph{Proceedings of ACM SIGCOMM Conference on Internet
  Measurement}, 2007.

\bibitem{TechCrunch15}
TechCrunch, ``2015 ad spend rises to \$187b, digital inches closer to one third
  of it,''
  \url{http://techcrunch.com/2015/01/20/2015-ad-spend-rises-to-187b-digital-inches-closer-to-one-third-of-it/},
  2015, [Online; accessed 19-October-2015].

\bibitem{Netflix16}
N.~Techblog, ``It’s all a/bout testing: The netflix experimentation
  platform,''
  \url{http://techblog.netflix.com/2016/04/its-all-about-testing-netflix.html},
  2016, [Online; accessed 10-March-2016].

\bibitem{Intelligence16}
Intelligence, ``Using dark posts to a/b test videos on facebook,''
  \url{http://intelligence.r29.com/post/130204487611/using-dark-posts-to-ab-test-videos-on-facebook},
  2016, [Online; accessed 10-March-2017].

\bibitem{Szabo10}
G.~Szabo and B.~A. Huberman, ``Predicting the popularity of online content,''
  \emph{Communications of the ACM}, vol.~53, no.~8, pp. 80--88, Aug. 2010.

\bibitem{Borghol11}
Y.~Borghol, S.~Mitra, S.~Ardon, N.~Carlsson, D.~L. Eager, and A.~Mahanti,
  ``Characterizing and modelling popularity of user-generated videos.''
  \emph{Performance Evaluation}, vol.~68, no.~11, pp. 1037--1055, 2011.

\bibitem{Bandari12}
\BIBentryALTinterwordspacing
R.~Bandari, S.~Asur, and B.~A. Huberman, ``{The Pulse of News in Social Media:
  Forecasting Popularity},'' \emph{CoRR}, vol. abs/1202.0332, 2012. [Online].
  Available: \url{http://arxiv.org/abs/1202.0332}
\BIBentrySTDinterwordspacing

\bibitem{Pinto13}
H.~Pinto, J.~M. Almeida, and M.~A. Gon\c{c}alves, ``Using early view patterns
  to predict the popularity of youtube videos,'' in \emph{{Proceedings of ACM
  International Conference on Web Search and Data Mining}}, 2013, pp. 365--374.

\bibitem{Khosla14}
A.~Khosla, A.~D. Sarma, and R.~Hamid, ``What makes an image popular?'' in
  \emph{Proceedings of International World Wide Web Conference (WWW)}, 2014.

\bibitem{Xu14}
J.~Xu, M.~van~der Schaar, J.~Liu, and H.~Li, ``Forecasting popularity of videos
  using social media,'' \emph{CoRR}, vol. abs/1403.5603, 2014.

\bibitem{Osborne11}
M.~Osborne and V.~Lavrenko, ``V.: Rt to win! predicting message propagation in
  twitter,'' in \emph{Proceedings of International Conference on Web and Social
  Media (ICWSM)}, 2011.

\bibitem{Hong11}
L.~Hong, O.~Dan, and B.~D. Davison, ``Predicting popular messages in twitter,''
  in \emph{Proceedings of International Conference Companion on World Wide
  Web}, 2011.

\bibitem{Gelli15}
F.~Gelli, T.~Uricchio, M.~Bertini, A.~D. Bimbo, and S.-F. Chang, ``Image
  popularity prediction in social media using sentiment and context features,''
  in \emph{Proceedings of the 23rd ACM International Conference on Multimedia},
  ser. MM '15, 2015.

\bibitem{Wu16b}
B.~Wu, W.-H. Cheng, Y.~Zhang, and T.~Mei, ``Time matters: Multi-scale
  temporalization of social media popularity,'' in \emph{Proceedings of the
  2016 ACM on Multimedia Conference}, ser. MM '16, 2016.

\bibitem{Wu16}
B.~Wu, T.~Mei, , and W.-H. C.~Y. Zhang, ``Unfolding temporal dynamics:
  Predicting social media popularity using multi-scale temporal
  decomposition,'' in \emph{Proceedings of the Thirtieth AAAI Conference on
  Artificial Intelligence}, ser. AAAI'16, 2016.

\bibitem{Almeida96}
V.~Almeida, A.~Bestavros, M.~Crovella, and A.~de~Oliveira, ``{Characterizing
  Reference Locality in the WWW},'' in \emph{{Proceedings of Conference on
  Parallel and Distributed Information Systems}}, 1996.

\bibitem{Chesire01}
M.~Chesire, A.~Wolman, G.~M. Voelker, and H.~M. Levy, ``Measurement and
  analysis of a streaming-media workload,'' in \emph{Proceedings of USENIX
  Symposium on Internet Technologies and Systems}, 2001.

\bibitem{Castillo14}
C.~Castillo, M.~El-Haddad, J.~Pfeffer, and M.~Stempeck, ``Characterizing the
  life cycle of online news stories using social media reactions,'' in
  \emph{Proceedings of ACM Conference on Computer Supported Cooperative Work
  And Social Computing}, 2014.

\bibitem{Tsagkias10}
M.~Tsagkias, W.~Weerkamp, and M.~de~Rijke, ``News comments: Exploring,
  modeling, and online prediction.'' in \emph{{Proceedings of European
  Conference on Information Retrieval}}, 2010.

\bibitem{Tatar14}
A.~Tatar, M.~D. de~Amorim, S.~Fdida, and P.~Antoniadis, ``A survey on
  predicting the popularity of web content,'' \emph{Journal of Internet
  Services and Applications}, vol.~5, 2014.

\bibitem{Jiang14}
L.~Jiang, Y.~Miao, Y.~Yang, Z.~Lan, and A.~G. Hauptmann, ``Viral video style: A
  closer look at viral videos on youtube,'' in \emph{{Proceedings of ACM
  International Conference on Multimedia Retrieval}}, 2014.

\bibitem{Crane08}
R.~Crane and D.~Sornette, ``Robust dynamic classes revealed by measuring the
  response function of a social system,'' \emph{Proceedings of National Academy
  of Sciences}, no.~41, pp. 15\,649--15\,653, 2008.

\bibitem{Drucker96}
H.~Drucker, C.~J.~C. Burges, L.~Kaufman, A.~J. Smola, and V.~Vapnik, ``Support
  vector regression machines,'' in \emph{Proceedings of Neural Information
  Processing Systems}, 1996, pp. 155--161.

\bibitem{Roy13}
S.~D. Roy, T.~Mei, W.~Zeng, and S.~Li, ``Towards cross-domain learning for
  social video popularity prediction.'' \emph{IEEE Transactions on Multimedia},
  vol.~15, no.~6, pp. 1255--1267, 2013.

\bibitem{Zhang16}
J.~Zhang, L.~Nie, X.~Wang, X.~He, X.~Huang, and T.~S. Chua,
  ``Shorter-is-better: Venue category estimation from micro-video,'' in
  \emph{Proceedings of the 2016 ACM on Multimedia Conference}, ser. MM '16,
  2016.

\bibitem{Chen16}
J.~Chen, X.~Song, L.~Nie, X.~Wang, H.~Zhang, and T.-S. Chua, ``Micro tells
  macro: Predicting the popularity of micro-videos via a transductive model,''
  in \emph{Proceedings of the 2016 ACM on Multimedia Conference}, ser. MM
  '16.\hskip 1em plus 0.5em minus 0.4em\relax ACM, 2016, pp. 898--907.

\bibitem{Viola01}
P.~A. Viola and M.~J. Jones, ``Rapid object detection using a boosted cascade
  of simple features,'' in \emph{{CVPR}}, 2001, pp. 511--518.

\bibitem{Jacobs04}
A.~Jacobs, A.~Miene, G.~T. Ioannidis, and O.~Herzog, ``Automatic shot boundary
  detection combining color, edge, and motion features of adjacent frames,'' in
  \emph{TRECVID 2004 Workshop Notebook Papers}, 2004, pp. 197--206.

\bibitem{Canny86}
J.~Canny, ``A computational approach to edge detection,'' \emph{{IEEE}
  Transactions on Pattern Analysis and Machine Intelligence}, vol.~8, no.~6,
  pp. 679--698, Jun. 1986.

\bibitem{Rosten10}
E.~Rosten, R.~Porter, and T.~Drummond, ``Faster and better: A machine learning
  approach to corner detection,'' \emph{{IEEE} Transations on Pattern Analysis
  and Machine Intelligence}, vol.~32, no.~1, pp. 105--119, 2010.

\bibitem{Calonder12}
M.~Calonder, V.~Lepetit, M.~{\"{O}}zuysal, T.~Trzcinski, C.~Strecha, and
  P.~Fua, ``{BRIEF:} computing a local binary descriptor very fast,''
  \emph{{IEEE} Transactions on Pattern Analysis and Machine Intelligence},
  vol.~34, no.~7, pp. 1281--1298, 2012.

\bibitem{He15}
K.~He, X.~Zhang, S.~Ren, and J.~Sun, ``Deep residual learning for image
  recognition,'' \emph{arXiv preprint arXiv:1512.03385}, 2015.

\bibitem{Hastie01}
T.~Hastie, R.~Tibshirani, and J.~Friedman, \emph{{The Elements of Statistical
  Learning}}, ser. Springer Series in Statistics.\hskip 1em plus 0.5em minus
  0.4em\relax Springer New York Inc., 2001.

\bibitem{Boujelbene10}
S.~Z. Boujelbene, D.~B.~A. Mezghanni, and N.~Ellouze, ``Improving svm by
  modifying kernel functions for speaker identification task,''
  \emph{International Journal of Digital Content Technology and its
  Applications}, vol.~4, no.~6, pp. 100--105, 2010.

\bibitem{Zhang07}
Z.~Zhang, R.~Min, and A.~Bonner, ``Modifying kernels using label information
  improves svm classification performance,'' in \emph{2007 International
  Conference on Machine Learning and Applications}, 2007.

\bibitem{Amari99}
S.~Amari and S.~Wu, ``Improving support vector machine classifiers by modifying
  kernel functions,'' \emph{Neural Networks}, vol.~12, no.~6, pp. 783--789,
  1999.

\bibitem{Figueiredo14}
F.~Figueiredo, J.~M. Almeida, M.~A. Gon\c{c}alves, and F.~Benevenuto, ``On the
  dynamics of social media popularity: A youtube case study,'' \emph{ACM
  Transactions on Internet Technology}, vol.~14, no.~4, pp. 24:1--24:23, Dec.
  2014.

\bibitem{Hofmann09}
T.~Hofmann, B.~Sch\"{o}lkopf, and A.~J. Smola, ``Kernel methods in machine
  learning,'' \emph{Annals of Statistics}, vol.~36, no.~3, pp. 1171--1220,
  2008.

\end{thebibliography}
% Generated by IEEEtran.bst, version: 1.13 (2008/09/30)

\begin{IEEEbiography}[{\includegraphics[width=1in,height=1.25in,clip,keepaspectratio]{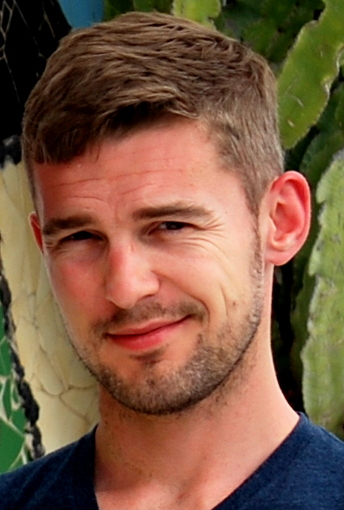}}]{Tomasz Trzci\'{n}ski} is an Assistant Professor in the Division of Computer Graphics in the Institute of Computer Science at Warsaw University of Technology since 2015. His main research interests include computer vision, machine learning and social media. He obtained his Ph.D. in Computer Vision at \'{E}cole Polytechnique F\'{e}d\'{e}rale de Lausanne in 2014. He received his M.Sc. degree in Research on Information and Communication Technologies from Universitat Polit\`{e}cnica de Catalunya and M.Sc. degree in Electronics Engineering from Politecnico di Torino in 2010. His professional appointments include work with Google, Qualcomm Corporate R\&D and Telefónica R\&D. In 2016, he was named a New Europe 100 Innovator as one of 100 outstanding challengers who are leading world-class innovation from Central and Eastern Europe. Since 2015, he holds a Chief Scientist position at Tooploox.
\end{IEEEbiography}

\begin{IEEEbiography}[{\includegraphics[width=1in,height=1.25in,clip,keepaspectratio]{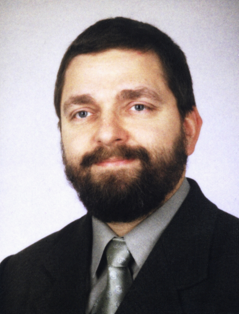}}]{Przemys\l{}aw Rokita}, MSc (1985), PhD (1993), DSc (2000), Tenured Professor (2014); Currently Professor and Head of the Division of Computer Graphics at the Warsaw University of Technology; Member of SPIE, ACM, IEEE; Main scientific interests: computer science and information technology, digital image processing, computer graphics, image perception; Previously affiliated as visiting scientist and professor at: the Max-Planck-Institut für Informatik - Computer Graphics Department (Germany), The University of Aizu (Japan), Hiroshima Institute of Technology (Japan), Hiroshima Prefectural University (Japan), Imperial College of Science, Technology and Medicine (United Kingdom); Member of Program Committees and reviewer for international scientific conferences and journals, including: IEEE Computer Graphics and Applications, The Visual Computer, Real-Time Imaging, Opto-Electronics Review, Journal of Imaging Science and Technology, IEEE Transactions on Circuits and Systems for Video Technology, IEEE Transactions on Multimedia, ACM Siggraph, Eurographics, High Performance Graphics; Expert and consultant at the Polish National Centre for Research and Development, National Science Centre, Ministry of Science and Higher Education; Laureate of the Golden Chalk Awards and title of best lecturer at the Faculty of Electronics anand Information Technology of the Warsaw University of Technology (2005 and 2006).
\end{IEEEbiography}

\end{document}